\documentclass[11pt]{article}
\usepackage[sc]{mathpazo}
\usepackage{booktabs}
\usepackage{graphicx}
\usepackage{epstopdf}
\usepackage{amsmath}
\usepackage{amssymb}
\usepackage{bm}
\usepackage{paralist}
\usepackage{amsfonts}
\usepackage{amsmath}
\usepackage{amsthm}
\usepackage[usenames]{color}
\usepackage{array}
\usepackage{mathrsfs}
\usepackage{epic,eepic,pstricks}
\usepackage{rotating}

\usepackage[
colorlinks=true,
linkcolor=blue,
urlcolor=blue,
filecolor=blue,
citecolor=red,
pdfstartview=FitV,
pdftitle={},
pdfauthor={},
pdfsubject={},
pdfkeywords={},
pdfpagemode=None,
bookmarksopen=true
]{hyperref}

\usepackage{epsfig}

\usepackage{hyperref}

\begin{document}
	\vspace{0.3cm}
	\begin{centering}		
		\textbf{\Large{Seismic Damage Assessment of Instrumented Wood-frame Buildings: A Case-study of  NEESWood Full-scale Shake Table Tests\footnote{This paper is an extended version of the paper titled "Element-by-element seismic damage diagnosis and prognosis in minimally instrumented wood-frame buildings" presented at Engineering Mechanics Institute Conference 2018 (MS23-Advanced deep learning based SHM) and participated in the EMI SHM and Control Committee Student Paper Competition at MIT, Boston, MA, May 29-June 1, 2018.}}}
		\vspace{0.8cm}
		
		{\large  Milad Roohi$^{1}$, Eric M. Hernandez$^2$, David Rosowsky$^3$ }
		
		\vspace{0.5cm}
		
		\begin{minipage}{.9\textwidth}\small
			\begin{center}
				{\it $^1$ Ph.D. Candidate, University of Vermont, Burlington, VT, USA}\\
				
				{\it $^2$ Associate Professor, University of Vermont, Burlington, VT,  USA}\\
				
				{\it $^3$ Provost and Senior Vice President; Professor, University of Vermont, Burlington, VT, USA}
				
				\vspace{0.5cm}
				
				{\tt mroohigh@uvm.edu, eric.hernandez@uvm.edu, david.rosowsky@uvm.edu}
				\\ $ \, $ \\

			\end{center}
		\end{minipage}
		\begin{abstract}
			The authors propose a methodology to perform seismic damage assessment of instrumented wood-frame buildings using response measurements. The proposed methodology employs a nonlinear model-based state observer that combines sparse acceleration measurements and a nonlinear structural model of a building to estimate the complete seismic response including displacements, velocity, acceleration and internal forces in all structural members. From the estimated seismic response and structural characteristics of each shear wall of the building, element-by-element seismic damage indices are computed and remaining useful life (pertaining to seismic effects) is predicted. The methodology is illustrated using measured data from the 2009 NEESWood Capstone full-scale shake table tests at the E-Defense facility in Japan.
		\end{abstract}
	\end{centering}
	\section{Introduction}
	Vibration measurements from structures can be used for a variety of purposes, including but not limited to model calibration, code verification, structural health monitoring, and vibration control. In the past two decades, researchers and engineers have recognized the importance of seismic instrumentation for quantitative and rapid performance assessment of buildings during and immediately following earthquakes \cite{porter2004real,celebi2004real}. Despite the immediate appeal, there are technical, logistical, and economic challenges associated with seismic instrumentation and performance assessment in the case of wood-frame buildings. Building instrumentation and its maintenance can be expensive, and budget constraints may not allow for exhaustive floor-by-floor or component-level instrumentation. These buildings exhibit a high degree of nonlinear behavior even during moderate ground motions, which makes it challenging to interpret the measured seismic response and provide a quantitative measure of the estimated level of damage caused by an earthquake. 
	
	One objective of this paper is to derive an extended model-based state observer (EMBO) that be used for interpretation of the measured seismic response from instrumented building structures. The EMBO is capable of combining a detailed nonlinear structural model and noise contaminated measured response of a structural system to estimate the complete dynamic time history response at all degrees of freedom of the model. This nonlinear state observer is an extension of the model-based state observer (MBO) that has been derived and used for state estimation in real-world structural systems that behave linear or mildly nonlinear under input motion \cite{Hernandez2013,Roohi2018}. 
	
	 Specifically, the aim of this paper is to propose a methodology  for estimation of element-by-element seismic damage indices in minimally instrumented wood-frame buildings. The methodology first employs the EMBO to estimate complete seismic response of an instrumented wood-frame building. The estimated response is then used as input to mechanics-based damage models to quantify element level damage indices and perform seismic damage diagnosis and prognosis of a wood-frame building. The proposed framework is verified and validated using simulated and real measured data from a six-story wood-frame instrumented building as part of the 2009 NEESWood Capstone building shake tests conducted at the E-Defense facility in Japan.
	
	The paper begins with a section describing the system of interest and measurement model. The following section presents the derivation of the proposed EMBO. This is followed by a section describing the 2009 NEESWood Capstone full-scale tests conducted at the E-Defense shake table in Japan. The numerical portion of the paper starts with verification of the proposed methodology using simulated measurements from a nonlinear model of the NEESWood Capstone building under input motions during the full-scale tests. Finally, the proposed methodology is validated using real seismic response measurements from the NEESWood Capstone tests.
	
	\section{System of Interest and Measurement Model}
	This paper focuses on typical building structures in which floor diaphragms can be assumed to be rigid for in-plane deformations. For this type of structures, the response to seismically induced ground acceleration can be accurately modeled by the following simultaneous set of non-linear differential equations     
	\begin{equation}
	\mathbf{M}\ddot{q}(t)+\mathbf{C}_{D}\dot{q}(t)+F_r(q(t))=-\mathbf{M}\mathbf{b}_1\ddot{u}_g(t)+\mathbf{b}_2w(t) 
	\label{system}
	\end{equation}
	The vector $q(t)\in \mathbb{R}^{n}$ represents the relative displacement (with respect to the ground) of all stories. For most buildings of interest, this results in three independent components per floor (two lateral displacements and a rotation about the vertical axis). The number of degrees of freedom is denoted as $n$, $\mathbf{M}=\mathbf{M}^T \in  \mathbb{R}^{n \times n}$ is the mass matrix, $\mathbf{C}_{D}=\mathbf{C}_{D}^T \in \mathbb{R}^{n\times n}$ is the damping matrix, $F_r(\cdot) $ is the resultant global restoring force vector which is obtained from the contribution of individual shear wall restoring forces to the global diaphragm coordinates. The matrix $\mathbf{b}_1 \in  \mathbb{R}^{n \times r}$ is the influence matrix of the $r$ ground acceleration time histories defined by the vector $\ddot{u}_g(t) \in  \mathbb{R}^{r}$. The matrix $\mathbf{b}_2 \in  \mathbb{R}^{n \times p}$ defines the spatial distribution the vector $w(t) \in \mathbb{R}^p$, which in the context of this paper represents the process noise generated by unmeasured excitations and/or modeling errors. 
	
	To analyze the system model in Equation \ref{system}, the equation is re-written in incremental form between $t$ and $t+\Delta(t)$ as follows
	\begin{equation}
	\mathbf{M}\Delta\ddot{q}+\mathbf{C}_{D}\Delta\dot{q}+\Delta F_r =-\mathbf{M}\mathbf{b}_1\Delta\ddot{u}_g+            \mathbf{b}_2\Delta w 
	\label{system}
	\end{equation}
	where $\Delta \cdot = \cdot(t+\Delta t)-\cdot(t)$ and  $\Delta F_r$ (the increment in the global restoring force) is given by
	\begin{equation}
	\Delta F_r = \mathbf{K}_T(t)\Delta q
	\end{equation}
	where $\mathbf{K}_T(t)$ is the global tangent stiffness matrix at time $t$, as the contributing stiffness of each SDoF shear wall in the global stiffness matrix is load (or displacement) history dependent. 
	
	This paper assumes that measurements $y(t)$ of the dynamic response of the structure consist in horizontal accelerations measured in three independent and non-intersecting directions. Vertical accelerations are typically also measured, however, this paper focuses only on horizontal acceleration measurements. {\color{black} The vector of $m$ acceleration measurements $y(t)$ is modeled as
		\begin{equation}
		y(t) = -\mathbf{c}_2\mathbf{M}^{-1}\left[\mathbf{C}_d\dot{q}(t)+F_r(q(t)) -\mathbf{b_2}w(t)\right]+\nu(t)
		\end{equation} 
		where 
		$\mathbf{c}_2 \in  \mathbb{R}^{m \times n}$ is a Boolean matrix that maps the DoFs to the measurements, and $\nu(t) \in \mathbb{R}^{m \times 1}$ is the measurement noise. 
		
		\section{Extended model-based state observer (EMBO)}
		A nonlinear state observer for state estimation in nonlinear systems can be written in state-space form as
		\begin{eqnarray}\label{nss}
		\dot{\hat{x}}(t) =\bm{f}(\hat{x}(t))+\textbf{G}(y(t)-\textbf{C}\hat{x}(t))
		\end{eqnarray}
		where ${\hat{x}}(t)$ denotes mean state estimate, \textbf{G} is a feedback gain, \textbf{C} is the measurement matrix and is the measurement. We expand nonlinear function $\bm{f}(.)$ using a Taylor series around the current estimate of the state vector and obtain a first-order approximation by dropping higher order terms of power series as follows
		\begin{equation}\label{LinObserver}
		\begin{aligned}
		\dot{\hat{x}}(t) &=\mathbf{A}_{\hat{x}(t)}\hat{x}(t)+\textbf{G}[y(t)-\textbf{C}\hat{x}(t)]\\
		&=(\mathbf{A}_{\hat{x}(t)}-\mathbf{GC})\hat{x}(t) + \mathbf{G}y(t)
		\end{aligned}
		\end{equation}
		We assume that estimates of velocity are equal to the derivative of the estimates of displacement and choose the upper partition of feedback gain to be zero and lower partition to be $\mathbf{M}^{-1}\mathbf{c}_{2}^{T}\mathbf{E}$, where matrix $\mathbf{E}$ is a matrix free to be selected in order to minimize the trace of the state error covariance and maps the DoFs to the measurements \cite{Hernandez2013}. This choice  of $\textbf{G}$ makes the observer realizable as a modified finite element model of the system with added grounded dampers and excited by corrective forces that are obtained from velocity measurements scaled by added damper values. In second-order form, the proposed extended model-based state observer (EMBO) estimate of displacement response is given by
		\begin{eqnarray}\label{NMBO}
		\mathbf{M}\ddot{\hat{q}}(t)+(\mathbf{C}_{D}+\mathbf{c}_{2}^{T}\mathbf{E}\mathbf{c}_{2})\dot{\hat{q}}(t)+\textbf{F}_r(\hat{q}(t))=\mathbf{c}_{2}^{T}\mathbf{E}y(t)
		\end{eqnarray}
		where $\hat{q}(t)$ is the time history of the estimated response at all DoFs of the model and $y(t)$ is noise contaminated velocity measurements obtained from acceleration measurements. The main advantage of the EMBO is that the nonlinear state observer can be implemented in advanced structural simulation software packages and this allows the state observer to be computationally efficient in propagation of the state estimate through nonlinear dynamics of a system.
		\subsection{Selection of feedback gain matrix}
		To determine the feedback gain matrix $\textbf{E}$ the objective function to be minimized is the trace of the estimation error covariance matrix. Since for a general nonlinear multi-variable case, a closed-form solution for the optimal matrix $\mathbf{E}$ has not been found, a numerical optimization algorithm is used. To derive the optimization objective function, Equation \ref{NMBO} is linearized as follows
		\begin{eqnarray}\label{LNMBO}
		\mathbf{M}\ddot{\hat{q}}(t)+(\mathbf{C}_{D}+\mathbf{c}_{2}^{T}\mathbf{E}\mathbf{c}_{2})\dot{\hat{q}}(t)+\mathbf{K}_0{\hat{q}}(t)=\mathbf{c}_{2}^{T}\mathbf{E}y(t)
		\end{eqnarray}
		where the elements of the stiffness matrix $\mathbf{K}_0$ are given by
		\begin{eqnarray}
		\mathbf{K}_{0_{i,j}}=\frac{\partial \mathbf{F}_{r,i}}{\partial q_j}|_{q_{j}=0}
		\end{eqnarray} 
		By defining the state error as $e=q-\hat{q}$, the expression for the state estimation error  is given by
		\begin{eqnarray}\label{LNMBO}
		\mathbf{M}\ddot{\hat{e}}(t)+(\mathbf{C}_{D}+\mathbf{c}_{2}^{T}\mathbf{E}\mathbf{c}_{2})\dot{\hat{e}}(t)+\mathbf{K}_0e(t)=b_2u(t)-\mathbf{c}_{2}^{T}\mathbf{E}\nu(t)
		\end{eqnarray}
		To derive an expression for the state error covariance, we take Fourier transforms of both sides of Equation \ref{LNMBO} and obtain the following
		\begin{eqnarray}\label{StateError}
		(\mathbf{M}\omega^2+(\mathbf{C}_{D}+\mathbf{c}_{2}^{T}\mathbf{E}\mathbf{c}_{2})i\omega+\mathbf{K}_0)e(\omega)=b_2u(\omega)-\mathbf{c}_{2}^{T}\mathbf{E}\nu(\omega)
		\end{eqnarray}
		with $\mathbf{H}_o$ defined as
		\begin{eqnarray}
		\mathbf{H}_o=\left(-\mathbf{M}\omega^2+\left(\mathbf{C}_D+\mathbf{c}_2^T\mathbf{Ec}_2\right)i\omega+\mathbf{K}_0\right)^{-1}
		\end{eqnarray}
		From Equation \ref{StateError}, the expression for the state error estimate in the frequency domain is
		\begin{eqnarray}
		e(\omega)=\mathbf{H}_o(b_2u(\omega)-\mathbf{c}^{T}_{2}\mathbf{E}\nu(\omega))
		\end{eqnarray}
		and the error spectral density matrix $\pmb{\Phi}_{ee}$ is given by
		\begin{eqnarray}\label{Phie}
		\pmb{\Phi}_{ee}(\omega)=\mathbf{H}_o\mathbf{b}_2\pmb{S}_{ww}(\omega)\mathbf{b}_2^{T}\mathbf{H}_o^*+\mathbf{H}_o\mathbf{c}_2^T\mathbf{E}\pmb{S}_{vv}(\omega)\mathbf{E}^T\mathbf{c}_2\mathbf{H}_o^*
		\end{eqnarray}
		where the matrices $\pmb{\Phi}_{ww}(\omega)$ and $\pmb{\Phi}_{vv}(\omega)$ are the power spectral density of the uncertain excitation on the system and measurement noise, respectively. To select the $\mathbf{E}$ matrix, the following optimization problem must be solved
		\begin{eqnarray}\label{J}
		\begin{aligned}
		& \underset{\mathbf{E}}{\text{minimize}}
		& & J=tr(\mathbf{P}) \\
		& \text{subject to}
		& &  \mathbf{E} \in \mathbb{R}^{+}
		\end{aligned}
		\end{eqnarray}
		where $\mathbf{P}$ is the displacement estimation error covariance matrix given by
		\begin{eqnarray}\label{P}
		\mathbf{P}= \mathbb{E}\left[(q(t)-\hat{q}(t))(q(t)-\hat{q}(t))^T\right]=\int_{-\infty}^{+\infty}\pmb{\Phi}_{ee}(\omega)d\omega
		\end{eqnarray}
		
		With this selection of the feedback matrix $\mathbf{E}$, the EMBO becomes a modified nonlinear model of the system with added grounded dampers obtained from a linearized model of the system in the measurement locations and excited by forces that are linear combinations of the measurements proportional to the added dampers.
		\begin{figure}[h!]
			\centering
			\includegraphics[width=0.9\linewidth]{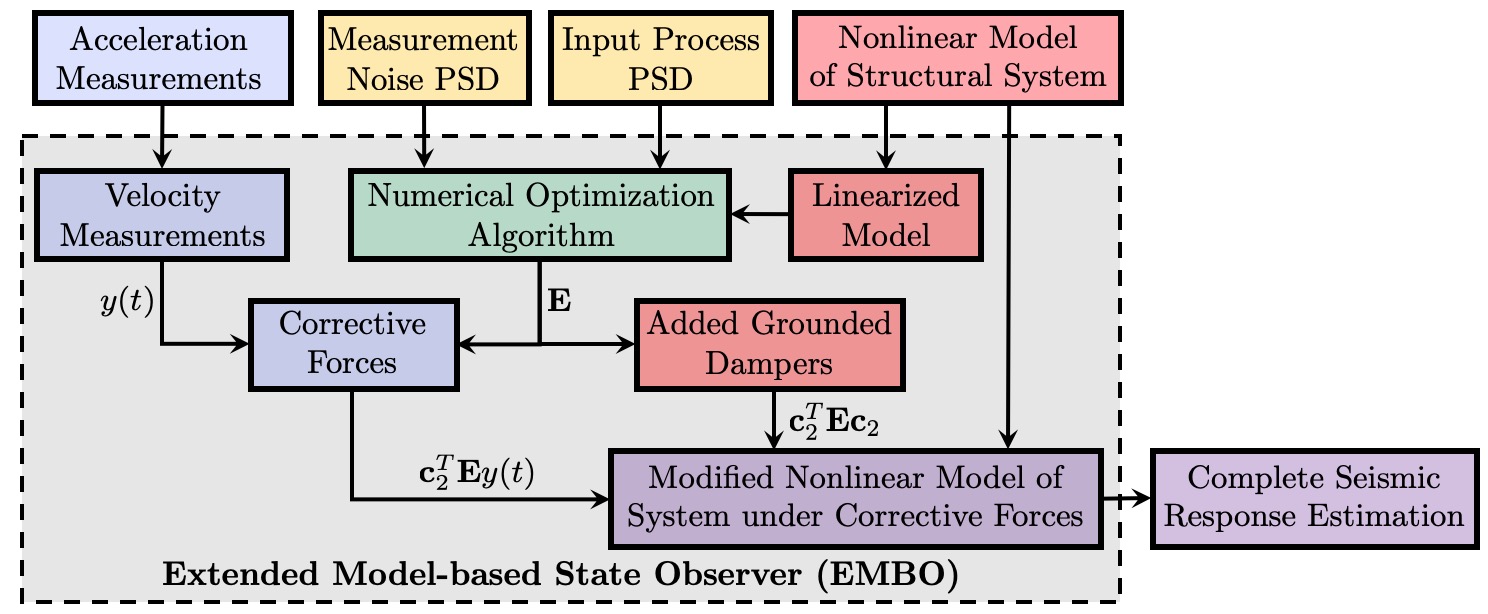}
			\vspace{-.4cm}
			\caption{Summary of the proposed EMBO for state estimation in nonlinear structural systems}
			\label{fig1}
		\end{figure}
		\section{Method of Approach}
		Our proposed methodology based on EMBO uses sparse response measurements, typically in the form of accelerations, to reconstruct the complete dynamic response of the structure. From the estimated response, a Park and Ang type low-cycle cumulative damage index can be constructed for every structural element in the building by \cite{Park1985}
		$$DI = \dfrac{\Delta_m}{\Delta_u}+\dfrac{\psi}{F_{ey}\Delta_u}\int{dE}$$
		where $\Delta_m$, $\Delta_u$ and $F_{ey}$ are estimated maximum deformation during the earthquake, ultimate deformation before collapse failure under monotonic loading determined experimentally and the equivalent yield force of the wall; and $\int{dE}$ is estimated incremental hysteretic energy absorbed by the wall during the earthquake. Also, $\psi$ is calibration parameter for the desired damage-based limit-state given by
		$$ \qquad 
		\psi = \beta_0 + \beta_1 x_{NS}^2 + \beta_2 x_{NS}^2 x_{WH}
		$$
		where $x_{NS}$ and $x_{WH}$ are nail spacing of the shear wall and width-to-height ratio of the shear walls; and $\beta_0$, $\beta_1$ and $\beta_2$ are regression coefficients calibrated from NEESWood full-scale shake table tests on a two-story light-frame wood building \cite{Park2009} which are presented in Table \ref{tab:addlabel}. Figure \ref{fig1} shows a summary of the methodology for element-by-element seismic damage index estimation.
		\vspace{-0.6cm}
		\begin{table}[htbp]
			\centering
			\caption{Regression coefficients calibrated from NEESWood full-scale shake table tests \cite{Park2009}}
			\vspace{0.3cm}
			\begin{tabular}{lrrr}
				\toprule
				Regression coefficient & \multicolumn{1}{l}{$\beta_1$} & \multicolumn{1}{l}{$\beta_2$} & \multicolumn{1}{l}{$\beta_3$} \\
				\midrule
				Value & 1.121 & 0.014 & 0.026 \\
				\bottomrule
			\end{tabular}%
			\label{tab:addlabel}%
		\end{table}%
		\begin{figure}[h!]
			\centering
			\includegraphics[width=0.33\linewidth]{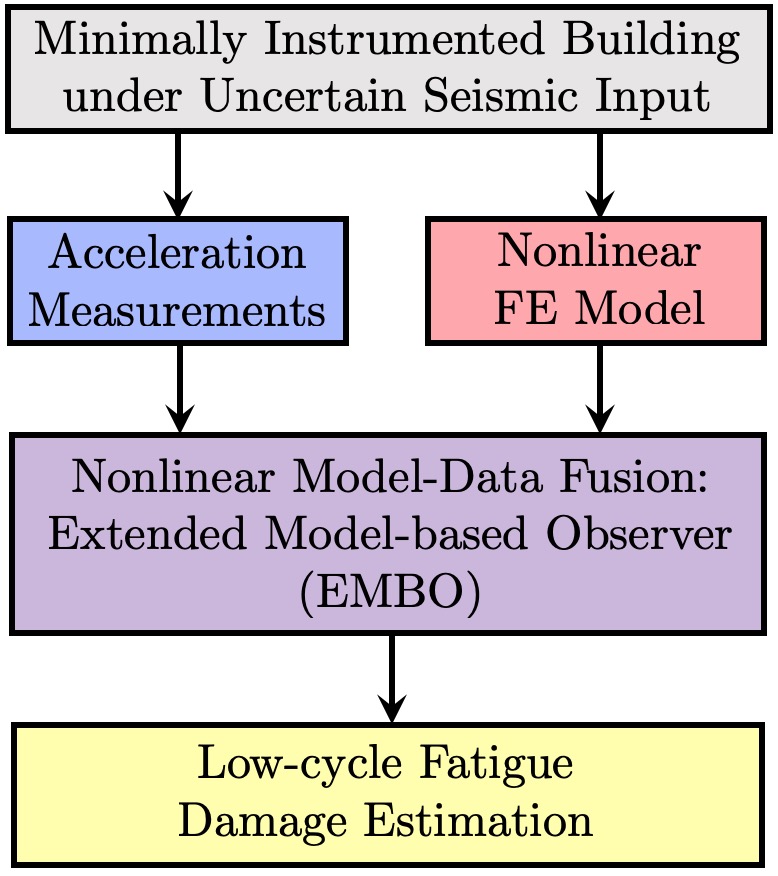}
			\vspace{-.4cm}
			\caption{Summary of the proposed methodology for element-by-element seismic damage index estimation using EMBO}
			\label{fig1}
		\end{figure}
		
		\section{Case study: NEESWood full-scale tests}
		We implemented the proposed methodology on a six-story wood-frame Capstone building tested in a series of full-scale seismic tests in the final phase of the NEESWood project. The building was tested with various hazard levels including (1) Test 3 (hazard level 50\% in 50 years), (2) Test 4 (hazard level 10\% in 50 years) and (3) Test 5 (hazard level 2\% in 50 years) and was instrumented with over 300 channels consisted of acceleration, displacement, strain and optical tracking measurements \cite{Lindt2010,Pang2010}. First, we verify the application of EMBO for state estimation using simulated response measurements. Then, we validate the proposed methodology using actual measured data from full-scale test of the building.
		
		\subsection{Verification using simulated response}
		In the verification step, a nonlinear 3D model of the building in OpenSEES (verified with M-SAWS model in \cite{Pang2010}) is used as a surrogate model and simulated data is generated by subjecting the model to the measured ground motion. The model includes every structural wall idealized as a pure shear element capable of resisting horizontal forces in its plane. The force-displacement relationship in each wall is modeled using the SAWS 10-parameter hysteretic model (Figure \ref{fig2}). 
		\begin{figure}[h!]
			\centering
			\includegraphics[width=0.9\linewidth]{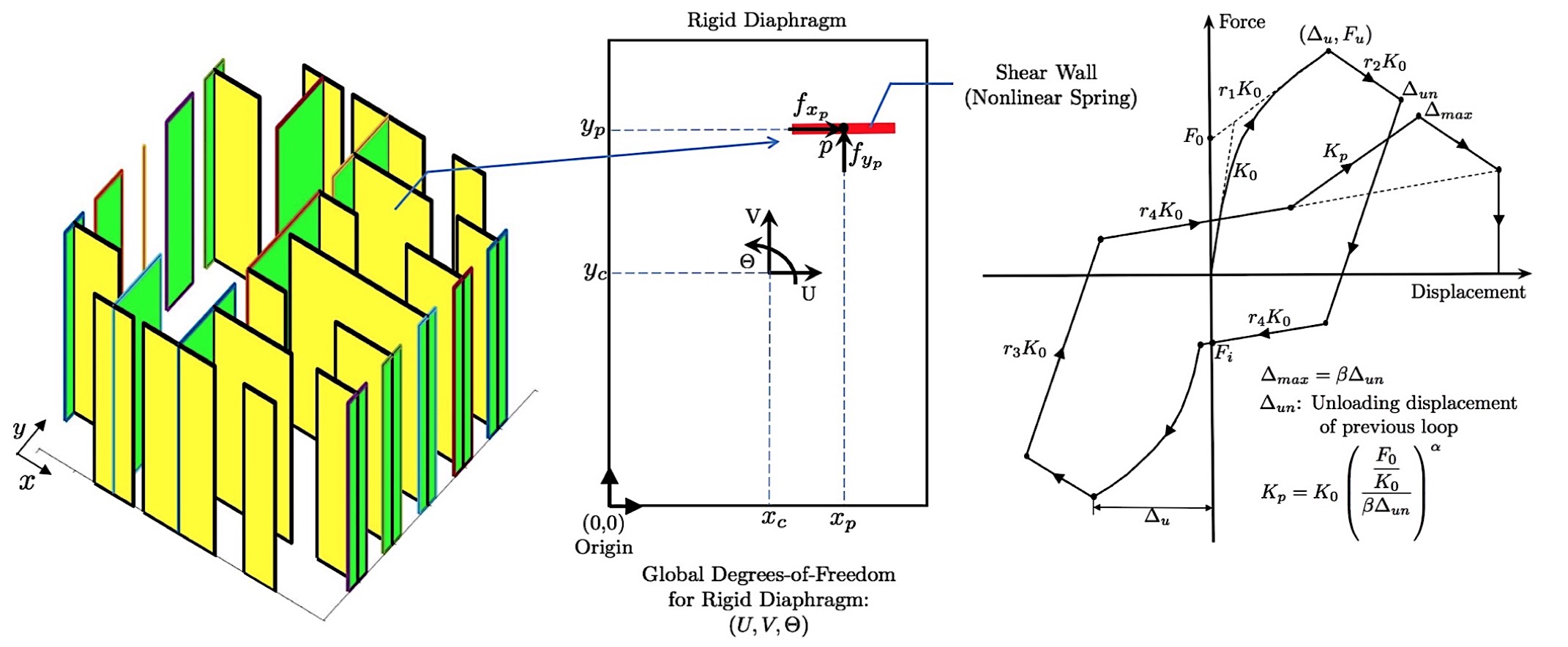}
			\vspace{-.4cm}
			\caption{Schematic of shear wall modeling using SAWS 10-parameter hysteretic model }
			\label{fig2}
		\end{figure}
		\begin{figure}[h!]
			\centering
			\includegraphics[width=0.9\linewidth]{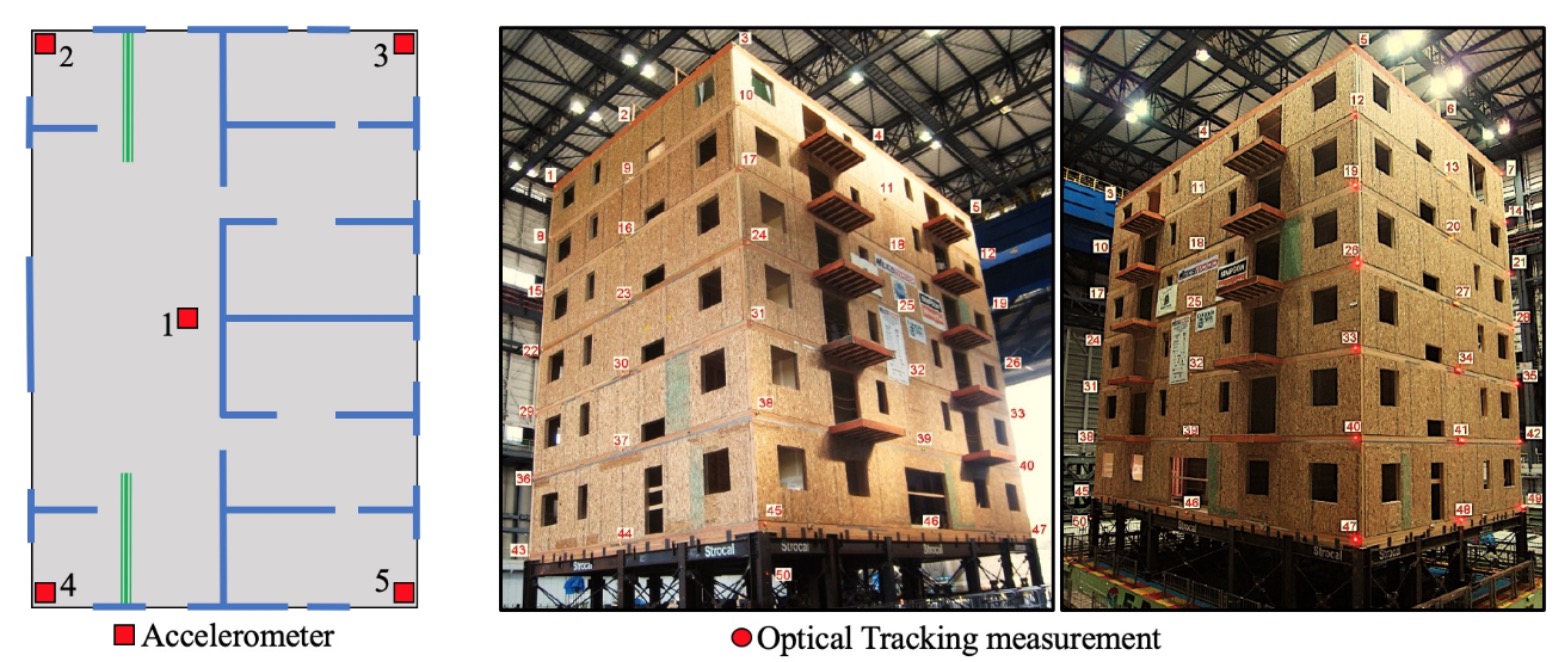}
			\vspace{-.4cm}
			\caption{Instrumentation locations: accelerometers in every floor (left) and optical tracking lasers (right)}
			\label{fig3}
		\end{figure}
		The OpenSEES model response during time history analysis under ground motion is assumed as the "real building response". The proposed EMBO model is implemented using a nonlinear 3D model of the building in OpenSEES with added grounded dampers at measured locations (Figure \ref{fig4}). We compared the observer estimates with the simulated building response and demands for every shear wall. Various measurement feedback scenarios were tested, and the results show that acceleration measurements at two floors (story three and roof with 3 measurements per floor) are enough to reconstruct the complete nonlinear dynamic response with high accuracy. The proposed observer provides very good tracking capabilities in demand estimates including nodal displacements, inter-story drifts and force-displacement hysteresis of shear walls using a relatively small number of measured seismic responses from the simulated building. Figure \ref{fig5} and Figure \ref{fig6} show a comparison of EMBO estimates of displacement and force-deformation for node 4 at story 5 (location 54 in Figure \ref{fig4}) with results from time history analysis under the ground motion from test 3.
		\begin{figure}[h!]
			\centering
			\includegraphics[width=0.9\linewidth]{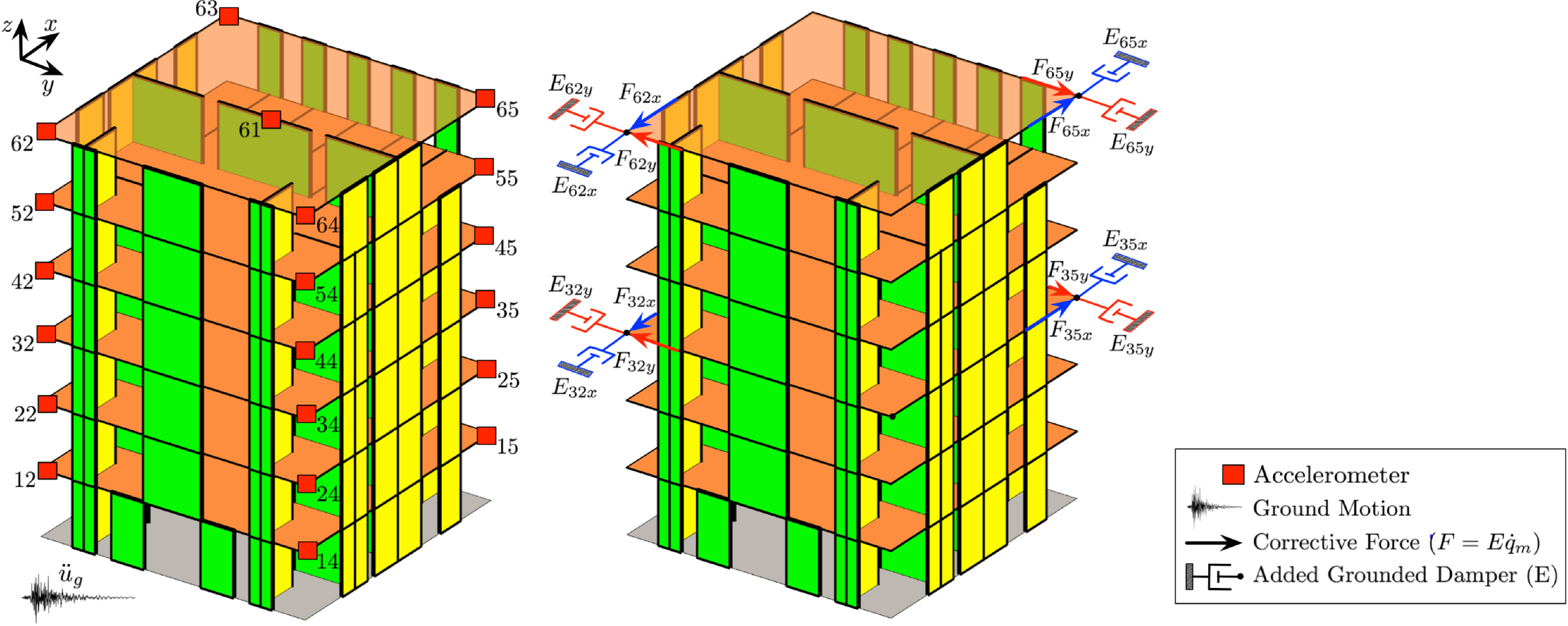}
			\vspace{-.4cm}
			\caption{OpenSEES Nonlinear 3D model with measurement locations (left) and OpenSEES EMBO model with added dampers in feedback locations and applied corrective forces (right)}
			\label{fig4}
		\end{figure}
		\begin{figure}[h!]
			\centering
			\includegraphics[width=0.35\linewidth]{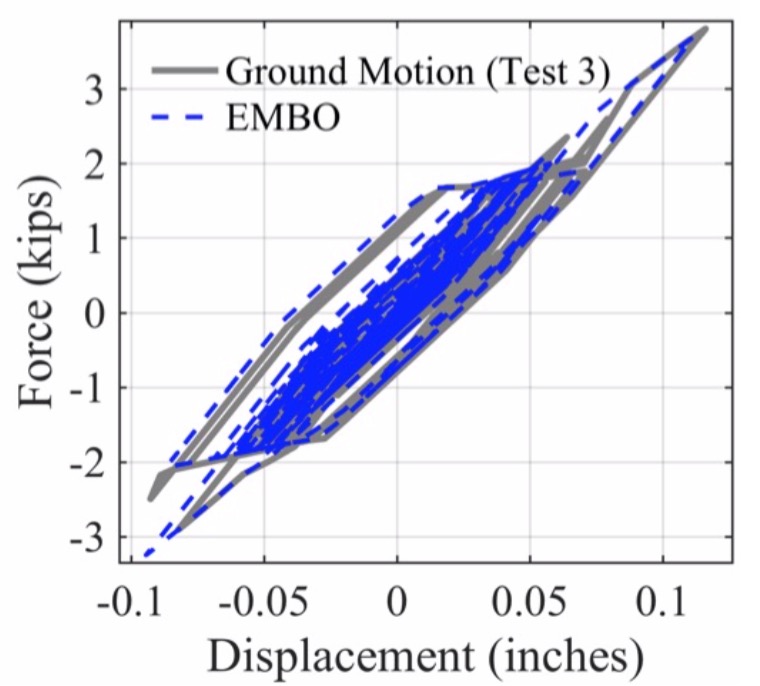}
			\vspace{-.4cm}
			\caption{Displacement comparison between EMBO estimation and time history analysis under ground motion (simulation)}
			\label{fig5}
		\end{figure}
		\begin{figure}[h!]
			\centering
			\includegraphics[width=0.55\linewidth]{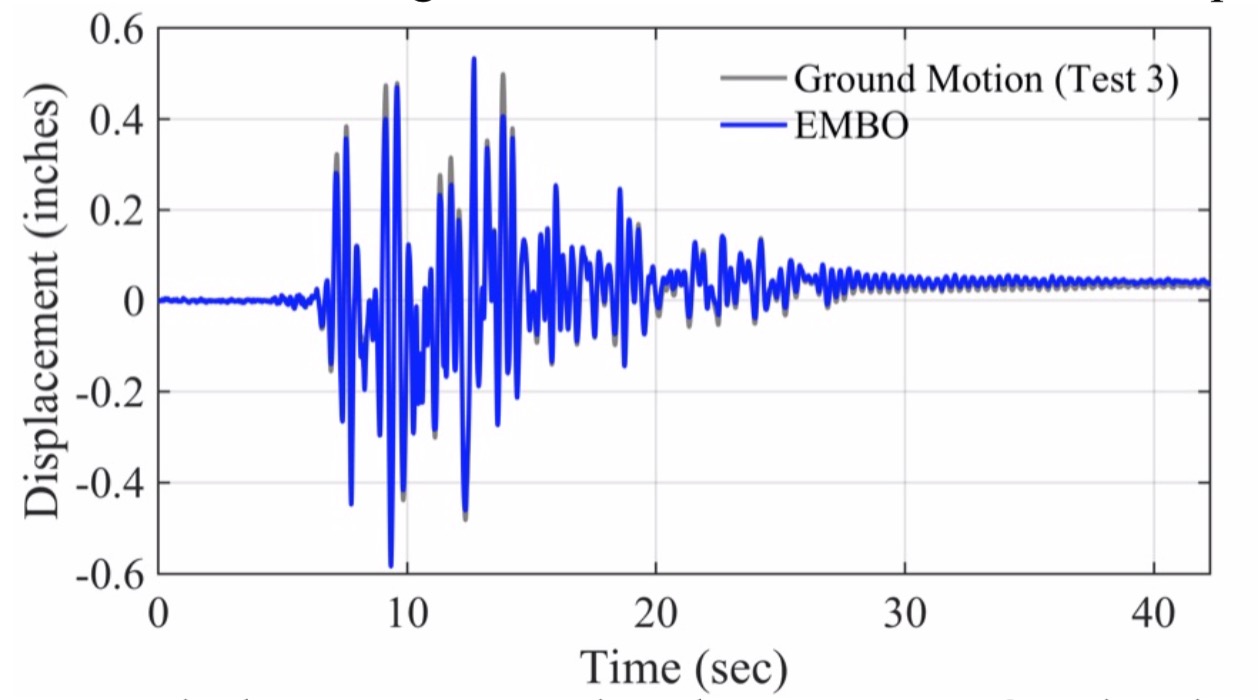}
			\vspace{-.4cm}
			\caption{Force-deformation comparison between EMBO and time history analysis under ground motion (simulation)}
			\label{fig6}
		\end{figure}
		\subsection{Validation using real data measured during shake table tests}
		In the validation case, we used measured data from the instrumented building as feedback to the same OpenSEES EMBO model as in the verification case. Complete demand estimates were computed for the structure and the results were compared with measured seismic responses from the tests. Figure \ref{fig7} and Figure \ref{fig8} show comparison of displacement and velocity estimates from EMBO and recorded data during test (location 54 in Figure \ref{fig4}).  Finally, from the estimated dynamic response and structural characteristics of each wall, an element-by-element damage index is computed. Figure \ref{fig9} shows estimated element-by-element damage indices at story 5 during various tests (hazard levels).
		\begin{figure}[h!]
			\centering
			\includegraphics[width=0.55\linewidth]{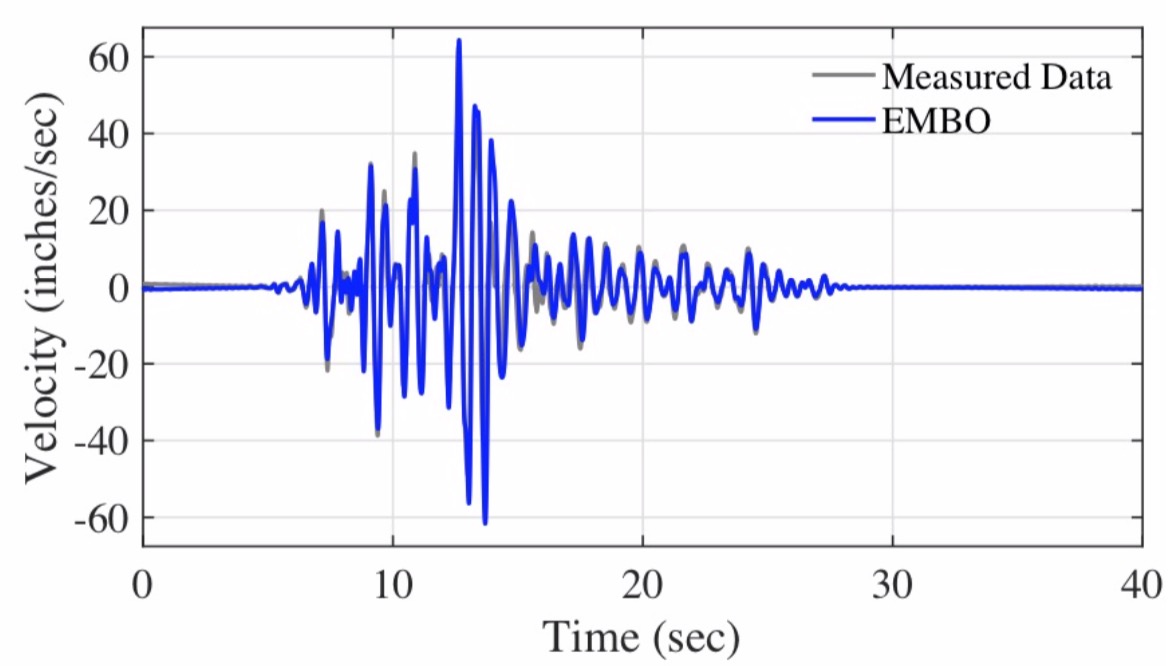}
			\vspace{-.4cm}
			\caption{Comparison of measured and EMBO estimated velocity at story 5 (location 54 in Figure \ref{fig4}) during test 5}
			\label{fig7}
		\end{figure}
		\begin{figure}[h!]
			\centering
			\includegraphics[width=0.55\linewidth]{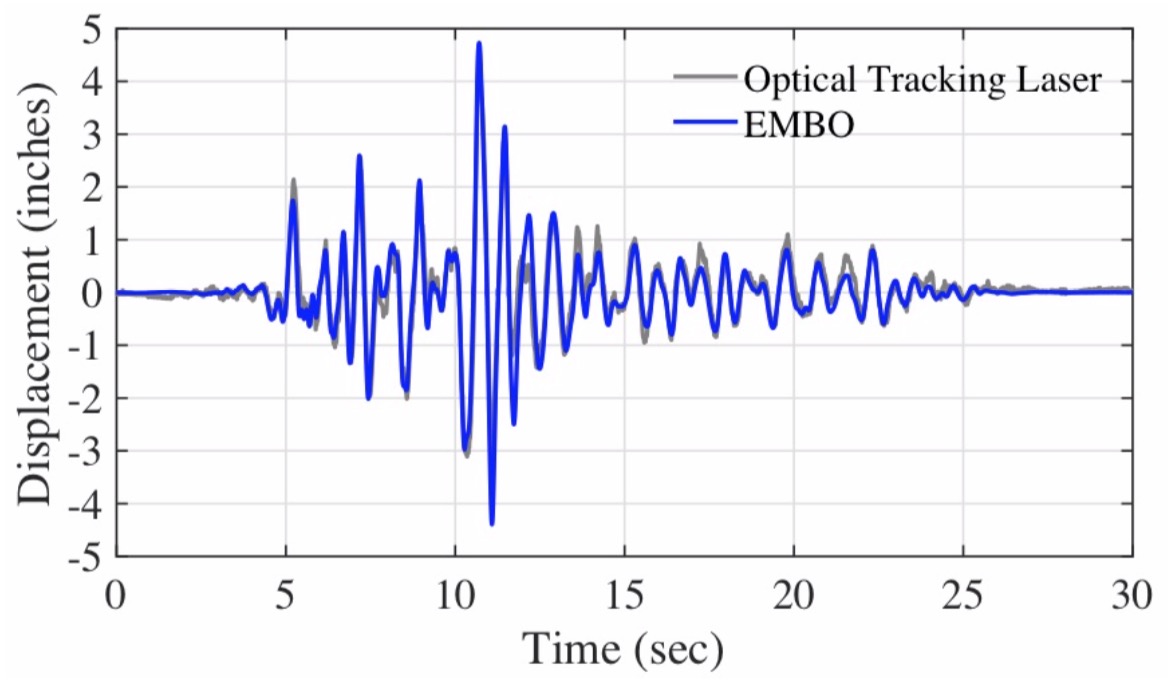}
			\vspace{-.4cm}
			\caption{Comparison of measured and EMBO estimated velocity at story 5 (location 54 in Figure \ref{fig4}) during test 5}
			\label{fig8}
		\end{figure}
		
		\begin{figure}[h!]
			\centering
			\includegraphics[width=0.7\linewidth]{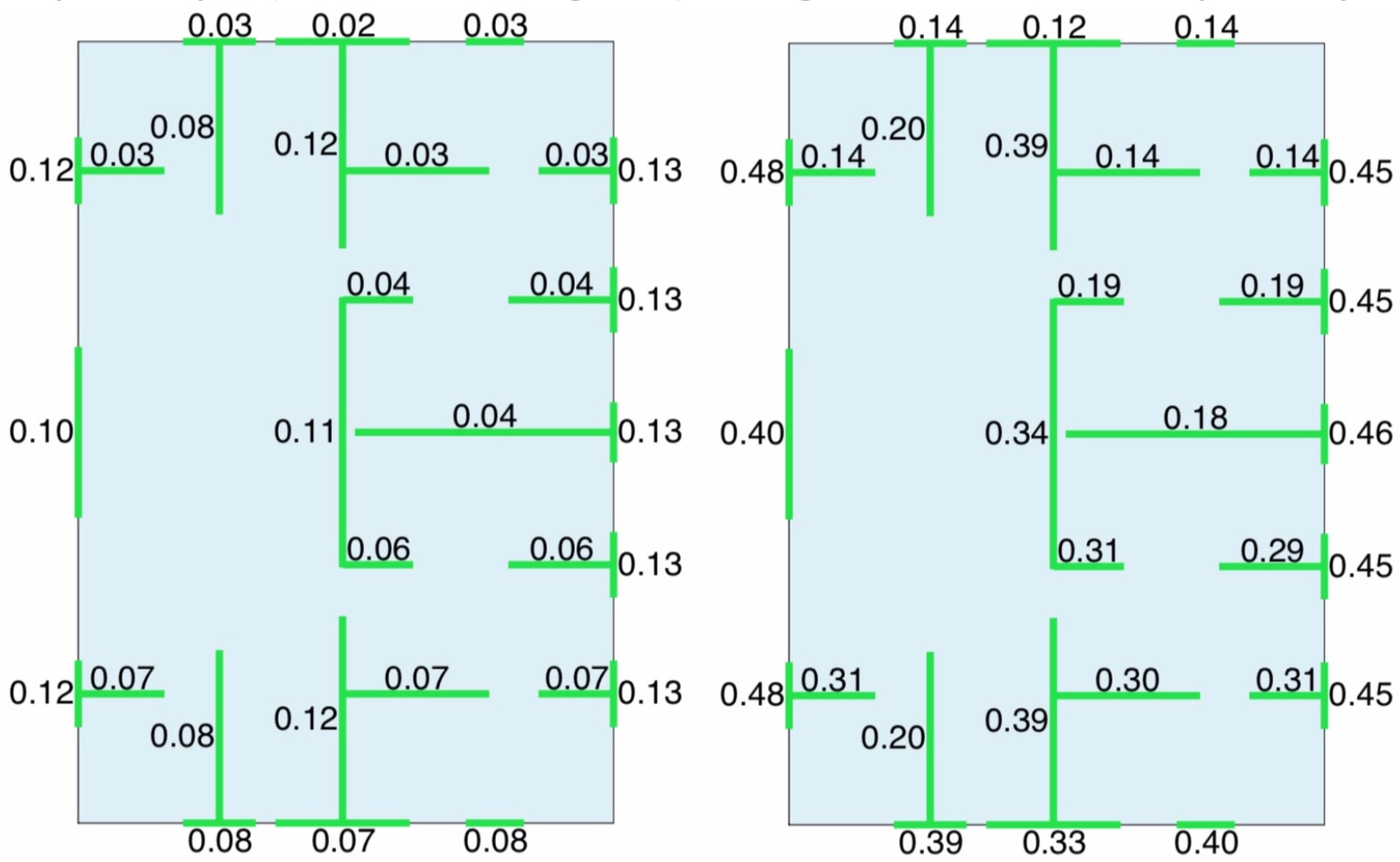}
			\vspace{-.4cm}
			\caption{Estimated element-by-element damage indices at story 5 during various tests: (left) test 3 and (right) test 4}
			\label{fig9}
		\end{figure}
		
		\section{Conclusion}
		This paper presents a methodology for element-by-element seismic damage diagnosis and prognosis of minimally instrumented wood-frame buildings. Seismic demands are computed using an extended model-based observer (EMBO). The EMBO combines a nonlinear structural model of the building and response measurements to reconstruct the complete dynamic response at all degrees of freedom of the model. The algorithm was successfully verified and validated using simulated and real data from a six-story wood-frame instrumented building as part of the NEESWood Capstone building shake test conducted at the E-Defense facility in Japan. Seismic damage index for every element of the structure was computed. Future work will focus on validating the estimated damage indices using pictures from the tests and on a methodology to assign uncertainty to the estimated responses and damage indices.

		\section{Acknowledgement}
		The authors would like to thank Professor John van de Lindt (Colorado State University) for his valuable support and his assistance with the acquisition and interpretation of the experimental data, Professor Bruce Ellingwood (Colorado State University) and Professor Shiling Pei (Colorado School of Mines) for the valuable discussion and suggestions. The authors would also like to thank Mr. Kazuki Tachibana (Tsukuba Research Institute) for his assistance in acquiring experimental data and photographic records. The research presented in this paper was partially funded by the National Science Foundation award No. 1453502. This support is gratefully acknowledged.


\begin{thebibliography}{}
			\bibitem{porter2004real}
			Porter, K., Mitrani‐Reiser, J. and Beck, J.L., 2006. Near-real-time loss estimation for instrumented buildings. The Structural Design of Tall and Special Buildings, 15(1), pp.3-20.
			\bibitem{celebi2004real}
			Celebi, M., Sanli, A., Sinclair, M., Gallant, S. and Radulescu, D., 2004. Real-time seismic monitoring needs of a building owner—and the solution: a cooperative effort. Earthquake Spectra, 20(2), pp.333-346.
			\bibitem{Hernandez2013} 
			Hernandez, E.M., 2013. Optimal model‐based state estimation in mechanical and structural systems. Structural control and health monitoring, 20(4), pp.532-543.
			\bibitem{Roohi2018} 
			Hernandez, E., Roohi, M. and Rosowsky, D., 2018. Estimation of element‐by‐element demand‐to‐capacity ratios in instrumented SMRF buildings using measured seismic response. Earthquake Engineering \& Structural Dynamics, 47(12), pp.2561-2578.
			\bibitem{Park1985}
			Park, Y.J., Ang, A.H.S. and Wen, Y.K., 1985. Seismic damage analysis of reinforced concrete buildings. Journal of Structural Engineering, 111(4), pp.740-757.
			\bibitem{Park2009}
			Park, S. and van de Lindt, J.W., 2009. Formulation of seismic fragilities for a wood-frame building based on visually determined damage indexes. Journal of Performance of Constructed Facilities, 23(5), pp.346-352.
			\bibitem{Pang2010}
			Pang, W., Rosowsky, D.V., Van de Lindt, J.W. and Pei, S., 2010. Simplified direct displacement design of six-story NEESWood capstone building and pre-test seismic performance assessment. MCEER.
			\bibitem{Lindt2010}    
			Van de Lindt, J.W., Pei, S., Pryor, S.E., Shimizu, H. and Isoda, H., 2010. Experimental seismic response of a full-scale six-story light-frame wood building. Journal of Structural Engineering, 136(10), pp.1262-1272.
		\end{thebibliography}
	\end{document}